\documentclass[aps,prl,amsmath,amssymb,twocolumn]{revtex4-1}

\usepackage{graphicx}% Include figure files
\DeclareGraphicsExtensions{.pdf,.png,.jpg}

\newcommand{\ket}[1]{| #1 \rangle}

\begin{document}

\title{Entangling Atomic Spins with a Strong Rydberg-Dressed Interaction}

\author{Y.-Y. Jau}
\author{A. M. Hankin}
\author{Tyler Keating}
\author{I. H. Deutsch}
\author{G. W. Biedermann}
\affiliation{Sandia National Laboratories, Albuquerque, New Mexico 87123, USA}%
\affiliation{Center for Quantum Information and Control (CQuIC), University of
New Mexico, Albuquerque NM 87131}
\date{\today}

\begin{abstract}
Controlling quantum entanglement between parts of a many-body system is the key to unlocking the power of quantum technologies such as quantum computation, high-precision sensing, and the simulation of many-body physics. The spin degrees of freedom of ultracold neutral atoms in their ground electronic state provide a natural platform for such applications thanks to their long coherence times and the ability to control them with magneto-optical fields. However, the creation of strong coherent coupling between spins has been challenging. Here we demonstrate a strong and tunable Rydberg-dressed interaction between spins of individually trapped cesium atoms with energy shifts of order 1 MHz in units of Planck's constant. This interaction leads to a ground-state spin-flip blockade, whereby simultaneous hyperfine spin flips of two atoms are inhibited due to their mutual interaction. We employ this spin-flip blockade to rapidly produce single-step Bell-state entanglement between two atoms with a fidelity $\geq$ 81(2)\%, excluding atom loss events, and $\geq$ 60(3)\% when loss is included.
\end{abstract}

\maketitle

Pristine quantum control of many-body correlations is fundamental to realizing the power of quantum information processors (QIP).  Steady progress has continued in various platforms ranging from solid state spintronics~\cite{Awschalom2013} and superconductors~\cite{Devoret2013,Barends2014} to nanophotonics~\cite{OBrien2009} and ultracold trapped atoms, both ionic~\cite{Brown2011,Monroe2013,Nigg2014} and neutral~\cite{Nelson2007,Zhang2010,Wilk2010}.  Cold neutral atoms are particularly attractive as the ability to create entanglement between atoms would allow for greatly increased precision of interferometers for applications in clocks~\cite{Appel2009,Bloom2014,Gil2014}, and force sensors~\cite{Dickerson2013,Parazzoli2012,Steffen2012}.  In addition, cold atoms provide a natural platform for quantum-simulation of condensed matter physics~\cite{Bloch2012,Glaetzle15} and scalable digital quantum computers~\cite{Brennen1999,Jaksch1999,Jaksch2000}.  Controlled entanglement of neutral atoms, however, has been challenging, particularly if one seeks tunable interactions that are strong, coherent, and long-range ($\sim\mu$m).

One mechanism to achieve strong, long-range coupling is the Rydberg blockade~\cite{Lukin1999}.  This has been successfully employed for implementing controlled entangling interactions between atoms~\cite{Wilk2010, Zhang2010, Hankin2014} and quantum logic gates~\cite{Isenhower2010}.  In the standard protocol, short pulses excite the population of one atom to the Rydberg state and optical excitation of a second atom is blockaded because of the electric dipole-dipole interaction (EDDI)~\cite{Jaksch2000}.  An alternative protocol is to adiabatically dress the ground state with the excited Rydberg state~\cite{Bouchoule2002, Johnson2010, Balewski2014}.
% Compared to directly utilizing Rydberg blockade,  -- Comment -- I don't think we should make this contrast.  We don't need to do it.
This Rydberg-dressed interaction enables tunable, anisotropic interactions that open the door to quantum simulations of a variety of novel quantum phases~\cite{Johnson2010,Pupillo2010, Yan2011}.  In addition, it allows for quantum control of interacting atoms based solely on microwave/rf fields whose phase coherence is easily maintained.  Applications include spin-squeezing for metrology \cite{Bouchoule2002,Gil2014},  and  quantum computing~\cite{Keating2013, Keating2014}.  While the promise of Rydberg-dressed interactions is great, experimental demonstration has been elusive.   We  present here a clear measurement of this interaction between two Rydberg-dressed atoms and employ coherent control in the ground state manifold to create entangled Bell-states.

\begin{figure}[htbp]
\includegraphics[scale=0.57]{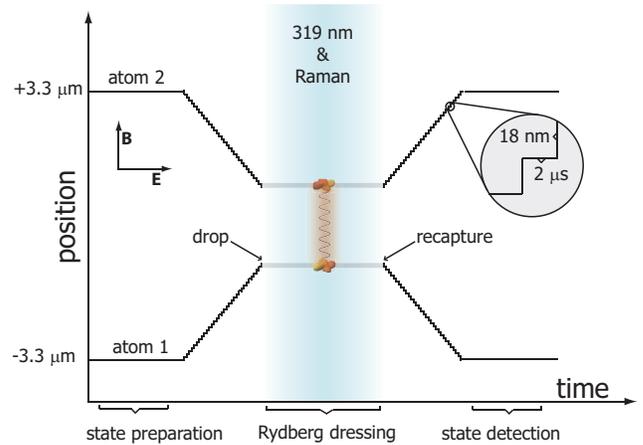}
\caption{\label{PD} \bf Experiment sequence. \rm To achieve both a strong ground-state atom-atom interaction and high-fidelity signal detection we perform the following steps at different interatomic spacings. In the experimental procedure, two Cs atoms are initially 6.6 $\mu$m apart, and held by optical tweezers. After qubit-state preparation, the two trapped atoms translate toward each other with an average speed of 9 mm/s (18 nm step every 2 $\mu$s) by ramping the modulation frequencies of the AOM.  At the target distance, the Rydberg dressing laser at 319 nm turns on to illuminate the two atoms simultaneously with a Raman laser. The tweezers are extinguished during this step to eliminate optical perturbation.  The two atoms then translate back to the original positions for state detection.}
\end{figure}

Rydberg-dressing arises from the modification of the light-shift (LS) (i.e., the optical AC Stark shift) of two atoms due to the EDDI.  We characterize this by the interaction strength $J \equiv \Delta E^{(2)}_{\rm LS} - 2 \Delta E^{(1)}_{\rm LS}$, where $ \Delta E^{(2)}_{\rm LS}$ is the LS for two atoms in the presence of EDDI and $\Delta E^{(1)}_{\rm LS}$ is the LS for a lone atom optically excited near a Rydberg level. By tuning the Rydberg excitation laser near the Rydberg resonance, double excitation of two Rydberg atoms is blockaded when $U_{\rm dd}\gg \hbar \Omega_{\rm L}$, where $\Omega_{\rm L}$ is the laser Rabi frequency and  $U_{\rm dd}$ is the EDDI energy, which scales from $1/R^6$ to $1/R^3$ depending on the interatomic distance $R$. The resulting difference in the two-atom spectrum implies that $\Delta E^{(2)}_{\rm LS}\neq2 \Delta E^{(1)}_{\rm LS}$, which plateaus at $J = \frac{\hbar}{2}\left[\Delta_L+\rm{sign}(\Delta_L)\left(\sqrt{\Delta^2_L+2\Omega^2_L}-2 \sqrt{\Delta^2_L+\Omega^2_L}\right)\right]$, where $\Delta_{\rm L}$ is the laser detuning, when the blockade is perfect~\cite{Johnson2010}. In the strong dressing regime, this can lead to a very strong interaction, $J\sim \hbar \Omega_L$, when $\Omega_L \sim \Delta_L$.

We employ $^{133}$Cs atoms, and encode qubits in the hyerfine ``clock states'' $\ket{0} \equiv \ket{6S_{1/2}, F=4, m_F = 0}$, $\ket{1} \equiv \ket{6S_{1/2}, F=3, m_F = 0}$, of two-atoms individually trapped in optical tweezers at 938 nm and separated by a few microns~\cite{Hankin2014}.  By choosing $|\Delta_{\rm L}|$ small compared to the 9.2 GHz hyperfine splitting, in the presence of EDDI, only the two-qubit state, $\ket{0,0}$, receives a nonnegligible, two-atom dressing energy $J$.  Because this energy is a shift of the Rydberg-dressed ground state, it can be used to control two-body interactions solely with mw-frequency fields.  In particular, by applying 9.2 GHz radiation on the clock-state resonance (e.g., via  two-photon Raman lasers) with Raman-Rabi frequency $\Omega_{\rm mw}$, two atoms initially in the state $\ket{1}$ are {\em blockaded} from undergoing double spin-flips, $\ket{1,1} \rightarrow \ket{0,0}$, when $J/\hbar \gg \Omega_{\rm mw}$.  This is a spin-flip blockade (Fig.~\ref{JvsR}a) in the dressed-ground-state, analogous to the familiar excited-state Rydberg blockade.

\begin{figure*}[t]
\includegraphics[trim=35 300 0 0,clip,scale=0.7]{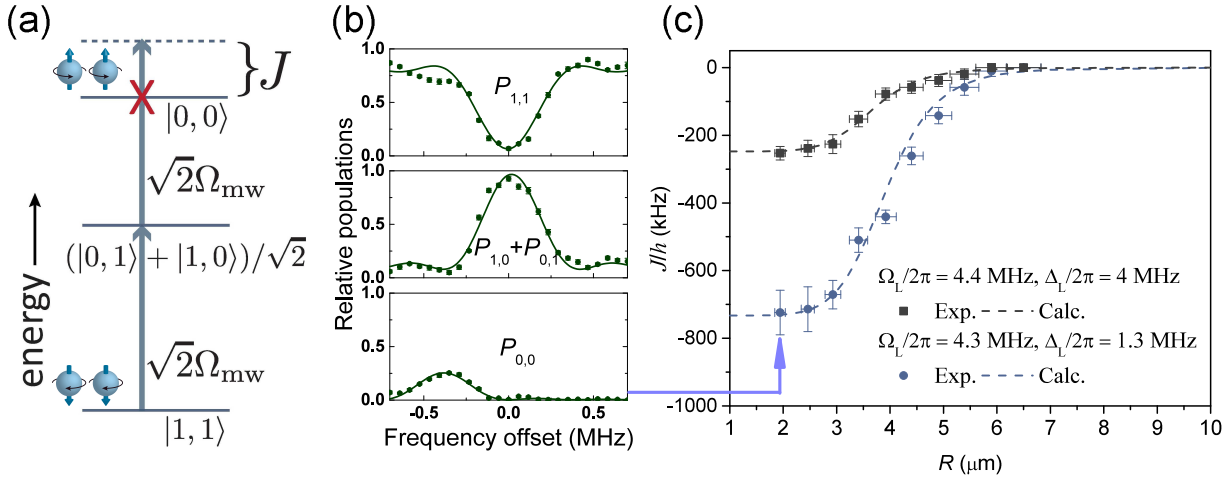}
\caption{\label{JvsR} \bf Rydberg-dressed ground-state interaction $J$ and the spin-flip blockade. \rm (a) Energy-level diagram of the spin-flip blockade on the Rydberg-dressed two-qubit sublevels: For a sufficiently large $J$, only the transition from $|1,1\rangle \rightarrow (|1,0\rangle$ or $|0,1\rangle)$ is allowed and the double spin-flip transition from $(|1,0\rangle$ or $|0,1\rangle) \rightarrow |0,0\rangle$ is blockaded when microwave radiation (stimulated Raman transition in our case) is applied at the noninteracting, single-atom qubit resonance frequency. (b) Scanning the microwave frequency of the stimulated Raman pulse applied to the two trapped Rydberg-dressed $^{133}$Cs atoms reveals the ground-state spin-flip blockade. The excitation from $|1,1\rangle \rightarrow |0,0\rangle$ occurs via an anti-blockade two-photon transition. $J/h$ is simply twice the shift of the resonance frequency for excitation to the state $|0,0\rangle$. (c) Experimental data of $J$ versus $R$ with two sets of parameters. The dashed curves are the calculated values based on a detailed model with no free parameters (see Supplementary Material). }
\end{figure*}

Note, unlike the single atom light-shift, the ratio of dressing energy $J$ to the photon scattering rate (the fundamental source of decoherence) improves {\em closer to resonance}, since far from resonance $J\sim\Delta^{-3}_L$, while absorption followed by spontaneous decay scales as $\gamma\sim\Delta^{-2}_L$.   For this reason, we operate in a regime of strong Rydberg dressing with a detuning, $\Delta_{\rm L}\leq\Omega_{\rm L}$.  For example, one of the experimental conditions used in Fig.~\ref{JvsR} is $\Omega_{\rm L}/2\pi=4.3$ MHz and $\Delta_{\rm L}/2\pi=1.3$ MHz, where the dressed state has a strong admixture of Rydberg character, $0.6|r\rangle+0.8|0\rangle$, or 64\% probability in $|0\rangle$.  Although there is about 36\% probability in $|r\rangle$ of the dressed state, the light shift on the Rydberg-dressed state is insensitive to thermal motion that gives rises to a fluctuation in the difference in the optical phase seen at the relative positions between the two atoms.  Such thermal noise was a  limiting factor in the generation of spin entanglement in the work of Wilk {\em et. al}~\cite{Wilk2010}. The thermal atomic motions in the Rydberg-dressed interactions, however, does lead to a Doppler shift, and thus noise in the optical detuning $\Delta_{\rm L}$, but this is highly suppressed when it is in the strong dressing regime ($\Delta_{\rm L}\leq\Omega_{\rm L}$), because $\Omega_{\rm L}$ dominates the energy shift of the dressed state.

The key mechanism that determines the interaction strength, $J$, depends on the EDDI and the optical Rabi frequency $\Omega_L$. Thus, the distance between the two Rydberg-dressed atoms and the choice of the principal quantum number are crucial experimental parameters. Ideally, we would like the atoms to be located far apart for individual addressing, and conversely, in close proximity to maximize $J$. Our particular implementation of an AOM (acousto-optic modulator) allows us to create two optical tweezers that trap each atom from the same laser by simultaneously driving the AOM at two frequencies (see Supplementary Material). We achieve this goal by independently sweeping the values of these frequencies and dynamically translating the traps. The capability of maximizing $J$ at shorter interatomic distance allows us to reduce the principal quantum number of the Rydberg level. Thus, the oscillator strength for direct excitation is improved, which allows us to maximize $\Omega_L$ for the same Rydberg laser intensity. This has the added benefit of reducing the interaction of the environment, which rapidly increases for high-lying Rydberg levels and is a common challenge in these experiments.

\section*{Results}

We directly measure $J$ as a function of the interatomic distance. Our experiment is illustrated in Fig.~\ref{PD}. The two trapped $^{133}$Cs atoms are initially prepared in state $\ket{1,1}$  and we dynamically translate them to be in close proximity at a targeted distance $R$. We then extinguish the tweezers for a short time to eliminate light shifts from the dipole-trap laser, and immediately apply a short pulse of the Raman and Rydberg-dressing lasers concurrently. Afterwards, the optical tweezers are restored to recapture the falling atoms and translate them back to the original positions for independent state detection, which is accomplished by using the $|6S_{1/2},F=4\rangle \rightarrow |6P_{3/2},F'=5\rangle$ $D2$ cycling transition to determine whether each atom is in state $|0\rangle$ (bright to this excitation) or $|1\rangle$ (dark to this excitation). We use a 319-nm laser for dressing the $^{133}$Cs atom, which couples atoms directly from the ground state to the Rydberg level,  $6S_{1/2}, F=4 \rightarrow 64P_{3/2}$, in a single photon transition~\cite{Hankin2014}. This avoids unwanted population in an intermediate, short-lived excited state that arises in the typical two-photon Rydberg excitation method, which causes additional ground-state decoherence~\cite{Keating2013}.  We choose a detuning that is small compared to the ground-state hyperfine splitting so that the dressing of  $\ket{1,1}$ in the $F=3$ manifold is negligible, but all other ground states in the logical basis, $\{\ket{0,0}, \ket{0,1}, \ket{1,0} \}$, are now well described in the dressed basis.  To drive spin flips, we apply the Raman laser fields to the two Rydberg-dressed atoms when they are at a desired separation.  By sweeping the Raman (microwave) frequency and measuring the resulting spin flips, we obtain a two-qubit energy spectrum of the form shown in Fig.~\ref{JvsR}b. To detemine $J$, we measure the microwave resonance frequency for the transition $|1,1\rangle \rightarrow (|1,0 \rangle +  |0,1 \rangle)/\sqrt{2}$ and for the two-microwave-photon transition $ |1,1 \rangle \rightarrow |0,0\rangle $.  $J/\hbar$ is given by twice the frequency difference of these two resonances. The microwave Rabi rate $\Omega_{\rm mw}$ and the pulse time of the stimulated Raman laser are properly chosen to ensure the observation of both the two-photon and the single-photon microwave resonance. The relative populations, $P_{1,1}$, $P_{1,0}$, $P_{0,1}$, and $P_{0,0}$ of the four, two-qubit computational basis states can be directly measured with the coincident bright and dark signals determined by the photon counting of the two APDs (avalanche photodiodes). Fig.~\ref{JvsR}c plots the measured $J$ as a function of interatomic distance for two different combinations of the 319-nm laser detuning $\Delta_{\rm L}$ and the optical Rabi rate $\Omega_{\rm L}$.

While the simple two-level atom model gives a clean theoretical prediction for $J$ as described above, the true atomic physics is more complex.   The EDDI shifts $64P_{3/2}$ negatively (red), thus for Rydberg-dressing we tune the 319-nm laser to the blue. At short interatomic spacings, however, the two-atom Rydberg energy levels strongly mix to yield a spectrum with molecular quality~\cite{Schwettmann2006}.  The resulting ``spaghetti'' of molecular levels could potentially lead to additional unwanted resonances that would ruin the Rydberg blockade. This may also affect the lifetime of the two-atom dressed state. The existence of such resonances, however, depends on the oscillator strengths. With our experimental resolution, we are unable to detect such resonances in our experiment. We use a best-fit of the blockade shift curve of $|r,r\rangle$ from our detailed model~\cite{Keating2013} to calculate $J$ as shown in Fig.~\ref{JvsR}c (see Supplementary Materials and Fig. S3).  With an ideal Rydberg blockade, $U_{\rm dd}\rightarrow\infty$, we find the expected plateau in $J$ at short interatomic distances predicted from the two-level model. The plateau at small $R$ is characteristic of a perfect Rydberg blockade, and agrees with simple theoretical predictions.

\begin{figure}[t]
\includegraphics[viewport=390 200 400 750, scale=0.36]{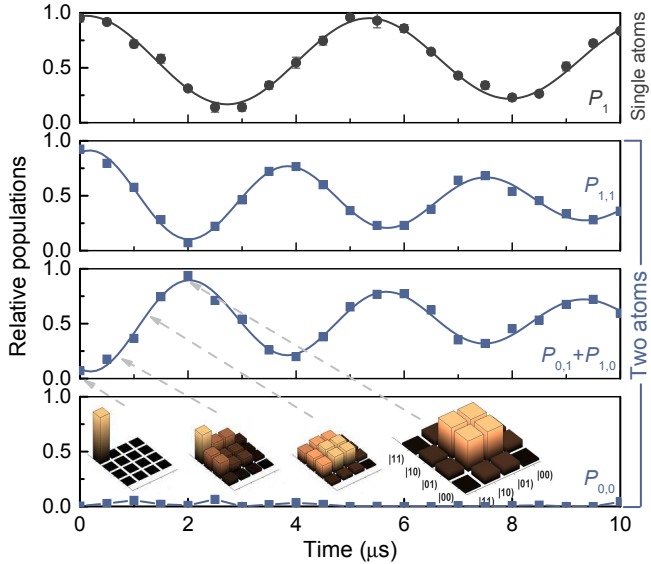}
\caption{\label{RDRO} \bf Generating entanglement directly. \rm Top panel: Rabi oscillations of a single Rydberg-dressed Cs qubit. Lower three panels: two-atom data with Rydberg-dressed spin-flip blockade ($J/h\approx750$ kHz with $\Omega_{\rm L}/2\pi=4.3$ MHz, $\Delta_{\rm L}/2\pi=1.1$ MHz, $R=2.9$ $\mu$m). The data points are fitted with curves of damped oscillation and exponentially varied offset. Rabi oscillation occurs between two spin down atoms and a two-qubit entangled state. There is a $\sqrt{2}$ enhancement of the microwave Rabi rate, $\Omega_{\rm{mw}}$ arising from the blockade, and excitation to state $|0,0\rangle$ is strongly suppressed due to the transition blockade as shown in Fig.~\ref{JvsR}a. The maximum Bell state $|\Psi_+\rangle$ entanglement is thus generated at around 2 $\mu$s.}
\end{figure}

With large values of $J$, we can employ the spin-flip blockade to create entanglement between atomic spin qubits. By driving a resonant Raman-pulse  $\ket{1} \rightarrow \ket{0}$ simultaneously on the two atoms, we cause Rabi oscillations between $\ket{1,1}$ and the entangled Bell state, $\ket{\Psi_+} = ( \ket{0,1} + \ket{1,0} )/\sqrt{2}$ as the experimental data show in Fig.~\ref{RDRO}. A signature of the spin-flip blockade is the characteristic increase in the Raman-Rabi frequency to $\sqrt{2} \Omega_{\rm mw}$~\cite{Wilk2010}.  The Bell state $\ket{\Phi_+} = ( \ket{0,0} + \ket{1,1} )/\sqrt{2}$, the two-atom cat state, can be generated from $\ket{\Psi_+}$ by subsequently applying a global $\pi/2$ rotation on the qubits. Alternatively, while the Rydberg dressing laser is still on,  a two-microwave-photon $\pi/2$-pulse at a shifted frequency, resonant with the transition from $|1,1\rangle \rightarrow |0,0\rangle$, also generates $\ket{\Phi_+}$.  This method has a lower fidelity because the two-photon microwave Rabi rate must be small in order to avoid excitation to the off-resonant $\ket{\Psi_+}$ state, and decoherence is more likely on this long time scale.

\begin{figure}[t]
\begin{center}
\includegraphics[viewport=370 230 400 710, scale=0.39]{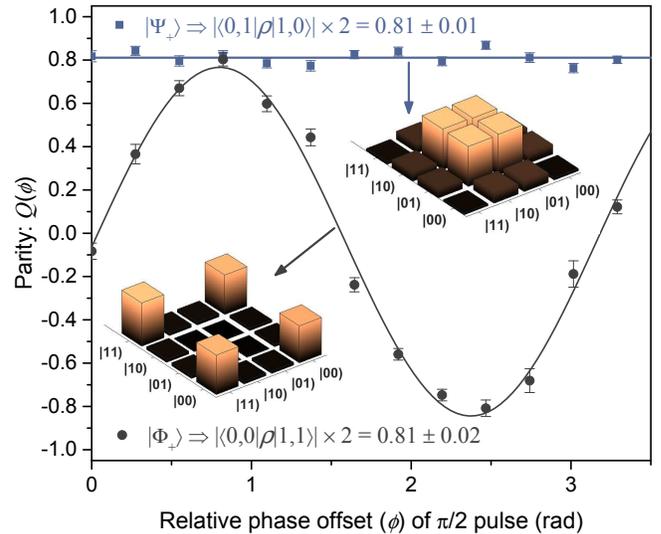}
\end{center}
\caption{\label{EW} \bf Entanglement verification. \rm A global $\pi/2$ pulse is applied to the undressed system after the entangled state is prepared. The data is obtained with the same experimental parameters used for data in Fig.~\ref{RDRO}. $\ket{\Psi_+}$ data is fitted with a straight line, and $\ket{\Phi_+}$ data is fitted with a sinusoidal function. It shows that both Bell states generated from our experiment have a fidelity $\geq$ 81(2)\%. Here, $\rho$ represents the two-qubit density matrix. The parity measurement, $P_{1,1} + P_{0,0} - (P_{0,1} + P_{1,0})$, allows direct determination of the amplitudes of the off-diagonal elements for both entangled states. }
\end{figure}

We measure the fidelity of Bell-state preparation as follows.  For a given $J$, the Raman pulse duration $T$ is chosen so that $\sqrt{2} \Omega_{\rm mw} T = \pi$ with $\Omega_{\rm mw}\ll J/\hbar$.  Following this procedure, the automated experimental control system checks whether both atoms are still present in the traps; if so, it counts as a ``valid'' operation. We determine the lower bound of the entanglement fidelity by measuring the off-diagonal coherence between the two-qubit logical basis states, $\langle x', y' | \rho | x, y \rangle$, where $\rho$ is the two-qubit density matrix.  For this, we apply a global $\pi/2$ pulse with phase $\phi$ to the entangled state.  As a function of $\phi$, we measure the expected value of the parity $Q(\phi) =\langle \sigma_z \otimes \sigma_z \rangle_\phi  = [P_{1,1}+ P_{0,0}- (P_{0,1} + P_{1,0})](\phi),$ where $P_{x,y}(\phi)$ is the population in the logical state $\ket{x,y}$ after application of the $\pi/2$ pulse~\cite{Bollinger1996, Sackett2000}.  For qubits prepared in the $\ket{\Psi_+}$ state, $Q$  is independent of $\phi$ and always a positive number; $Q$ = 1 when the entanglement is perfect. For qubits prepared in the $\ket{\Phi_+}$ state, $Q(\phi)$ is an oscillating function of $\phi$. In this case, perfect entanglement corresponds to perfect oscillation visibility. The entanglement fidelity (fidelity between the prepared state and the target Bell state) is $\ge 2| \langle x', y' | \rho | x, y \rangle |$, as measured from $Q(\phi)$~\cite{Sackett2000}.  When this fidelity is greater than 50\%, the state is necessarily entangled. The measurement of $Q(\phi)$ in Fig.~\ref{EW} shows that with the same experimental parameters as used in Fig.~\ref{RDRO}, we have achieved entanglement fidelity  $\geq$ $81\pm2$\% for generating both $\ket{\Psi_+}$ and $\ket{\Phi_+}$ when both atoms are recaptured in the trap.

\section*{Discussion}
Given that we measure the two-atom survival probability after the procedure to be $74\pm2$\%, the success rate of deterministically generating an entangled qubit pair is therefore $\ge 60\pm3$\%. With our current experimental data rate ($\approx$ 10 s$^{-1}$), on average we generate 6 pairs of entangled qubits per second. The current  factors limiting the entanglement fidelity given a valid procedure are: the optical pumping efficiency, decay of the Raman Rabi oscillation of the Rydberg-dressed states (Fig.~\ref{RDRO}), and the strength of $J$. Optical pumping efficiency determines how well we can prepare the atoms in the qubit subspace (computational space), which can be improved with a more careful pumping scheme. Decay of the Rabi oscillation impacts the fidelity of the $\pi$ pulse;  we can improve this fidelity by increasing $\Omega_{\rm mw}$ but only for sufficiently large $J$.  One identified contribution to the decay is the shorter than expected single-atom Rydberg-state lifetime (see Supplementary Material).  A comprehensive explanation of the two-atom Rabi oscillation decay is the subject of future work.

The protocol presented here is quite robust despite some technical imperfections, and we expect improvement in future experiments. For example, currently the Bell-state entanglement is generated in the dressed basis, resulting in atom loss due to the admixture of the Rydberg state.  In principle dressed states can be returned to the bare ground states by adiabatically ramping the dressing laser's intensity and detuning in an optimal way \cite{Keating2014}, something we will implement in a future generation of the experiment.  Finally, minimizing other experimental imperfections will likely improve the Rydberg state lifetime, improve the deterministic entanglement fidelity, and open the door to more full control of complex quantum systems.

\section{Supplementary Materials}

\subsection{Experimental platform}
\begin{figure}[b]
\center
\includegraphics[width=7 cm]{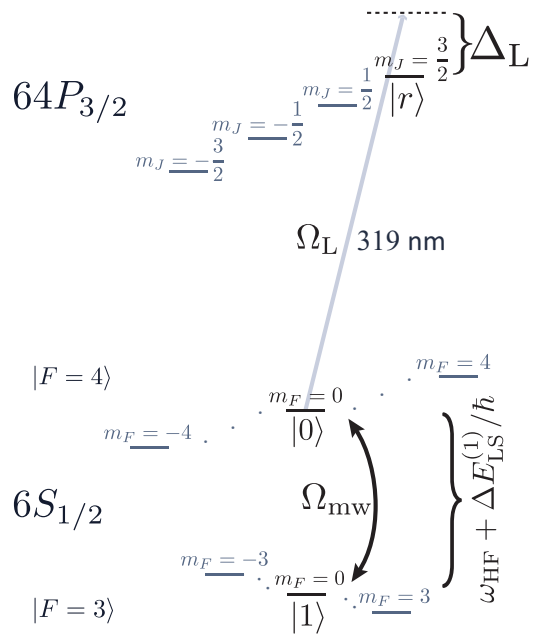}
\caption{\label{ELS} Relevant energy level diagram for the $^{133}$Cs atom.  The qubit states $|0\rangle$ and $|1\rangle$ are encoded in the clock state of the Cs hyperfine sublevels $|F=4,m_F=0\rangle$ and $|F=3,m_F=0\rangle$. The Rydberg dressing laser, detuned $\Delta_{\rm{L}}$ from $|64P_{3/2},m_J=3/2 \rangle$, strongly interacts with the $|F=4\rangle$ hyperfine manifold. Here, $\omega_{\rm HF} + \Delta E^{(1)}_{\rm LS}/\hbar$ is the hyperfine splitting summed with the single-atom light shift due to the 319-nm laser.  A two-photon stimulated Raman transition drives Rabi oscillations between $|0 \rangle$ and $|1\rangle$ at a rate $\Omega_{\rm{mw}}$.}
\end{figure}

Our experimental platform \cite{Hankin2014} is built on laser cooled $^{133}$Cs atoms, singly trapped in optical tweezers (dipole traps). We encode qubits in the clock states of the cesium $6S_{1/2}$ ground-state manifold with a hyperfine splitting $\omega_{HF}/2\pi=9.2$ GHz. We choose as our logical basis $|0\rangle = |6S_{1/2}, F=4,m_F=0\rangle$ and $|1\rangle = |6S_{1/2}, F=3,m_F=0\rangle$ as shown in Fig.~\ref{ELS}.  The two atoms are loaded from a MOT (magneto-optical-trap) into the two optical dipole traps with a separation of 6.6 $\mu$m. These dipole traps are generated by sending a 938-nm laser beam to an AOM (acousto-optic modulator), which is simultaneously modulated at two different frequencies. This creates two beams with a well-defined angular separation that depends linearly on the AOM drive frequency difference.  These beams pass through a focusing lens forming two tightly-focused spots with waists of $r_{1/e^2}$ = 1.29(3) $\mu$m and a well-defined spatial separation that depends on the angular separation of the two beams.  Following polarization gradient cooling, the atom temperature is reduced to $\approx$ 20 $\mu$K. A bias field at 4.8 G then turns on, and we optically pump the atoms into state $|6S_{1/2},  F=4,m_F=0\rangle$ (logical basis state $|0\rangle$) using a $\pi$-polarized laser at 895 nm tuned to $|6S_{1/2}, F=4\rangle \rightarrow |6P_{1/2}, F'=4\rangle$ and a repump laser at 852 nm tuned to $|6S_{1/2}, F=3\rangle \rightarrow |6P_{3/2}, F'=4\rangle$. The state preparation efficiency is $\approx$ 95\%, limited by the stray, fictitious magnetic field produced by vector light shifts from the dipole-trap laser. We apply a two-photon Raman laser field to perform a global $\pi$ rotation to bring the atoms from $|0,0\rangle \rightarrow |1,1\rangle$. The stimulated Raman transition uses the carrier and one sideband from a laser tuned 50-GHz red of the Cs $D2$ line ($6S_{1/2} \rightarrow 6P_{3/2}$) and modulated via a fiber-based EOM (electro-optic modulator).

To Rydberg-dress the atoms with a strong EDDI (electric dipole-dipole interaction), we dynamically translate the two Cs atoms into close proximity to a targeted distance $R$ by ramping the AOM modulation frequencies as shown in Fig.~1 of the main article. From the initial separation of 6.6~$\mu$m, we can continuously vary $R$ down to a minimum value of 1.5~$\mu$m, at which point the two traps begin to merge causing atom loss. During the Rydberg dressing period, we turn off the traps to eliminate the light shift due to the trap laser. Afterward, we restore the traps to recapture the falling atoms. We use a Rydberg dressing laser at 319 nm which drives direct, single-photon transitions from $6S_{1/2}$ to $nP_{3/2}$. This Rydberg excitation laser is designed to cover the principal quantum numbers ranging from $n=30$ to ionization. Because EDDI causes a red shift of the two-atom Rydberg state $|r,r\rangle$, we blue detune the Rydberg excitation laser. With a strong bias magnetic field, we use the Rydberg state $nP_{3/2}$ with magnetic sublevel $m_J=3/2$ for dressing the qubit state $|0\rangle$ (Fig.~\ref{ELS}).

For state detection, we translate the two trapped Cs atoms back to the original positions. The state-dependent detection is accomplished using the $|6S_{1/2},F=4\rangle \rightarrow |6P_{3/2},F'=5\rangle$ $D2$ cycling transition to determine whether each atom is in state $|0\rangle$ (bright to this excitation) or $|1\rangle$ (dark to this excitation). In the case that the atom is found to be dark, we immediately apply the repump laser simultaneously with the cycling laser to check that the atom is indeed in state $|1\rangle$ by verifying its presence in the trap.  Note that the detection method identifies the entire $|6S_{1/2}, F=4\rangle$ manifold with $|0\rangle$ and the entire $| 6S_{1/2}, F=3\rangle$ manifold with $|1\rangle$. The fluorescence signals are detected via two APDs (avalanche photodiodes) that are coupled via optical fiber. This non-destructive method allows us to reuse atoms without reloading new atoms from the MOT, increasing our data rate to $\approx$~10 s$^{-1}$ from $\sim1$~s$^{-1}$. The experimental protocol is carried out with an FPGA (Field Programmable Gate Array) control system~\cite{Hankin2014}.

\subsection{Generating Bell states}
In order to obtain a strong differential light shift, and therefore a strong dressed-interaction energy $J$ between two atoms, we require $\Delta_{\rm L}\leq\Omega_{\rm L}$ (the strong dressing regime). In fact, this condition also leads to a maximum suppression of the optical detuning noise due to the thermal motion of the atoms. The probability amplitude of the ground-state $|0\rangle$ in the dressed state is $(\Delta_{\rm L}/\sqrt{\Delta_{\rm L}^2+\Omega_{\rm L}^2}+1)^{1/2}/\sqrt{2}$, and has a value between $1/\sqrt{2}$ for maximal dressing and 1 for no dressing at all. In our experiment we choose two example conditions: ($\Omega_{\rm L}/2\pi=4.4$ MHz, $\Delta_{\rm L}/2\pi=4$ MHz) and ($\Omega_{\rm L}/2\pi=4.3$ MHz, $\Delta_{\rm L}/2\pi=1.3$ MHz). This leads to the dressed states of $0.41|r\rangle+0.91|0\rangle$ and $0.6|r\rangle+0.8|0\rangle$ or 84\% and 64\% probability in $|0\rangle$ respectively. In this experiment we do not adiabatically transfer the dressed state back to the bare ground state, which lowers the atom recapture probability. A detailed discussion regarding strong Rydberg dressing and adiabatic transfer can be found in Refs.\cite{Keating2013,Keating2014}. When there is a probability that the atom is excited to the Rydberg state, there is a probability that the atom will not be recaptured in the trap during the time window ($\approx$ 10 $\mu$s) of efficient recapture . In the detection, we only count the data with atoms that remain trapped. We track statistics concerning the lost atoms.

A straightforward method for generating the Bell state $|\Psi_+\rangle = (|0,1\rangle + |1,0\rangle)/\sqrt{2}$ is illustrated in Fig.~3 of the main article. This shows typical experimental data representing Rabi flopping in the presence of Rydberg dressing with the parameters $n=64$, $\Omega_{\rm L}/2\pi=4.3$ MHz, $\Delta_{\rm L}/2\pi=1.1$ MHz, $R=2.9$ $\mu$m, and $J/h \approx 750$ kHz. Each data point is the average of several hundred measurements but with various Raman pulse durations. With a strong $J$, microwave excitation to the state $|0,0\rangle$ is blockaded. The microwave Rabi oscillation can only occur between $|1,1\rangle$ and $(|0,1\rangle + |1,0\rangle)/\sqrt{2}$ with a very small probability exciting to $|0,0\rangle$. We also find a factor of $\sqrt{2}$ enhancement of the microwave Rabi rate compared to the single-atom Rabi rate as secondary evidence of entanglement.

In Fig.~3 of the main article, the optimal $\pi$ time for generating $|\Psi_+\rangle$ is $\approx$ 2 $\mu$s. To produce a Bell state $|\Phi_+\rangle$, we simply apply a global $\pi/2$ pulse to $|\Psi_+\rangle$. By fine tuning the experimental parameters, our best entanglement fidelity is $\geq$~81(2)\% excluding the atom loss events, and  $\geq$~60(3)\% when loss is included. If the atom recapture process were to completely filter all cases when the atom was excited to the Rydberg state, we would expect to have a lower two-atom survival probability compared to the 74\% recapture probabilty we present here. With perfect filtering and excluding the atom loss events, the entanglement fidelity would be much higher and eventually limited by the state preparation efficiency. However, it would still be near 60\% with loss included. For technical reasons, we observe that a fraction of the atoms that are excited to the Rydberg state decay back to the ground state within the recapture time window. This is only consistent with a shorter than expected Rydberg lifetime.  The resulting increase in the two-atom survival probability reduces the value of the fidelity measurement that is conditional on the atom loss events. We anticipate a substantial improvement of the entanglement fidelity by achieving close to the natural Rydberg-state lifetime and also by using an adiabatic ramp of the intensity and detuning of the 319 nm Rydberg laser in order to return the dressed state back to the bare ground state.

\subsection{Calculation of Rydberg-state interactions and the ground-state interaction $J$}
\begin{figure*}[t]
\includegraphics[width=17 cm]{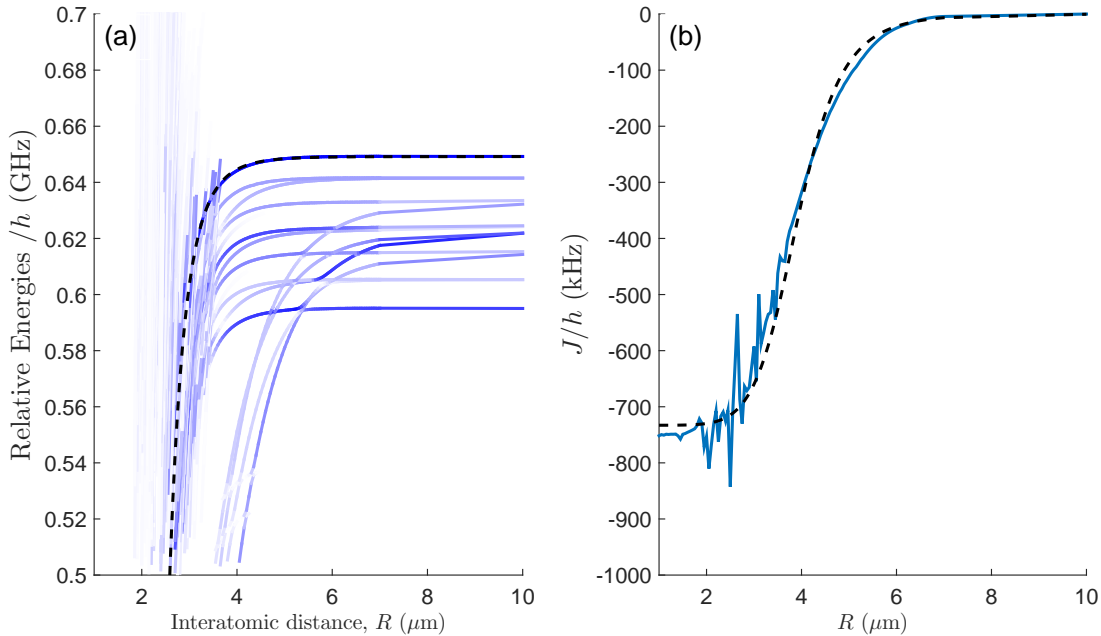}
\caption{\label{JvsR}(a) Calculated two-atom Rydberg sublevels of $64P_{3/2}$ with $|\bf E| \rm =6.4$ V/m and $|\bf B| \rm =4.8$ G. The vertical axis is the energy offset from the center of gravity of $64P$. The black dashed curve is the fitting result of our selected two-atom Rydberg state $|m_J=3/2, m_J=3/2\rangle$. (b) Solid curve: the calculated $J$ as a function of $R$ using all levels from the left panel. The parameters of the dressing laser are $\Omega_{\rm L}/2\pi$ = 4.3 MHz and $\Delta_{\rm L}/2\pi=1.3$ MHz. Dashed curve: using the black dashed curve from the left panel with the same parameters.}
\end{figure*}
Calculating $J$ as a function of the interatomic distance $R$ requires knowledge of how the two-atom Rydberg energy levels shift as a function of $R$. The mixing of atomic orbitals due to interactions leads to molecular-like energy levels. We use the same numerical codes we developed in our previous work~\cite{Keating2013,Hankin2014} for the calculation. Figure~\ref{JvsR}a shows the calculated two-atom molecular energy levels for $64P_{3/2}$ for $R$ from 1 $\mu$m to 10 $\mu$m with electric field $|\bf E| \rm =6.4$ V/m and magnetic field $|\bf B| \rm =4.8$ G,  parameters typical of our experiment. There are a total of 14,400 energy levels included in the calculation, which cover the principal quantum numbers from $n=60$ to $n=66$ and the orbitals from $l=0$ to $l=4$. In the calculation, the Rydberg-Rydberg interactions include EDDI, electric dipole-quadrupole, and electric quadrupole-quadrupole interactions.  As one can see, there is a ``spaghetti'' structure at very short $R$ due to the strong mixing between all combinations of Rydberg energy levels. When excited with the 319-nm laser, the probability of exciting one of these resonances in the spaghetti depends on the oscillator strength of the different molecular levels.  In the figure, we denote this oscillator strength by the darkness of the lines. We see that most molecular resonances are only weakly coupled to the Rydberg excitation laser. With limited computational resources, it is numerically intractable to calculate the entirety of multipole interactions in the infinite atomic basis set. Hence, the calculated spaghetti feature can never be precise. In Fig.~\ref{JvsR}b, we plot the two calculated $J(R)$ curves with parameters $\Omega_{\rm L}/2\pi=4.3$ MHz and $\Delta_{\rm L}/2\pi=1.3$ MHz using the result from Fig.~\ref{JvsR}a. The $J(R)$ represented in the solid curve uses all the energy levels we calculate. It has a wiggling structure at small $R$ mainly due to the multiple level crossing and mixing of the doubly-excited Rydberg levels. The $J(R)$ represented in the dashed curve is a best-fit curve for the doubly-excited Rydberg state of our interest ($|m_J=3/2, m_J=3/2 \rangle$) to represent $|r,r\rangle$ as a function of $R$ shown by the black-dashed curve in Fig.~\ref{JvsR}a. This simple blockade shift curve uses a form of $C/R^6$. The single fitting parameter $C$ is determined by using the all-level calculated data with $R\geq3.5$ $\mu$m. One can see that the two results of $J(R)$ in Fig.~\ref{JvsR}b are very similar aside from the wiggling structure on the solid curve. As mentioned before, this structure changes with different atomic basis sets used in the calculation, and we have not been able to obtain a converged answer with our limited computer power. We therefore use the dashed $J(R)$ curve to compare with our experimental data.

\subsection{Rydberg State Lifetime}
Decay of the microwave Rabi oscillation of the Rydberg dressed atoms is partially explained by the shorter than expected lifetime of the Rydberg-state. The longest measured Rydberg-state lifetime for a single atom in this experiment is on the order of 40 $\mu$s while the expected natural lifetime due solely to vacuum and blackbody radiation is $\approx$ 150 $\mu$s. Our measurements indicate that the Rydberg state lifetime is further reduced by photo-induced processes on the nearby ITO-coated surfaces, which are for shielding the external electric-field.  We performed direct measurements of the single-atom Rydberg state lifetime by using two resonant optical Rydberg excitation $\pi$ pulses with a variable delay time in between. By measuring the probability of the atom in the ground state following this sequence, we can determine the Rydberg state lifetime. We found that the Rydberg state lifetime was greatly reduced when the laser light acts on the nearby ITO coated surface. A study of this effect was carried out with a non-resonant 638-nm laser, which was originally used for generating charges on the ITO surfaces to compensate the background electric fields \cite{Hankin2014}. The experiment indicates that stronger 638-nm laser intensity leads to a shorter Rydberg state lifetime. We suspect that scattered laser light impinging on the nearby surface, such as the 319-nm and 938-nm lasers, can produce a similar effect. The longest lifetime was measured under the conditions of minimizing all possible laser powers into the vacuum chamber. In the situation of the entanglement experiment, the Rydberg-state lifetime is only about 10 $\mu$s. However, the detailed mechanism driving the reduction in the Rydberg state lifetime is the subject of further study.

\bibliography{RDB_arXiv}

\end{document}